\documentclass[aps,pre,showpacs,onecolumn]{revtex4}
\usepackage{bm}
\usepackage{graphicx}
\bibstyle{approve.bib}
\usepackage{amssymb}
\usepackage{amsmath}
\usepackage{esint}
\usepackage{epstopdf}
\usepackage{color}
\usepackage{gensymb}
\usepackage{mathtools}
\usepackage{suffix}

\newcommand{\be}{\begin{equation}}
\newcommand{\ee}{\end{equation}}
\newcommand{\bea}{\begin{eqnarray}}
\newcommand{\eea}{\end{eqnarray}}

\newcommand{\br}{\mathbf{r}}

\newcommand{\bE}{\mathbf{E}}
\newcommand{\bF}{\mathbf{F}}

\newcommand{\bb}{\mathbf{B}}

\newcommand{\lm}{L_{\rm m}}

\newcommand{\e}{\varepsilon}

\newcommand{\p}{_{\rm p}}

\newcommand{\ce}{_{\rm c}}

\newcommand{\ew}{\varepsilon_{\rm w}}

\begin{document}

\title{Sensing Translocating Polymers via Induced Magnetic Fields}           
\author{Sahin Buyukdagli\footnote{email:~\texttt{buyukdagli@fen.bilkent.edu.tr}}}
\affiliation{Department of Physics, Bilkent University, Ankara 06800, Turkey}
\date{\today}

\begin{abstract} The requirement to boost the resolution of nanopore-based biosequencing devices necessitates the integration of novel biosensing techniques with reduced sensitivity to background noise. In this article, we probe the signatures of translocating polymers in magnetic fields induced by ionic currents through membrane nanopores.  Within the framework of a previously introduced charge transport theory, we evaluate the magnetic field signals generated by voltage- and pressure-driven DNA translocation events in monovalent salt solutions. Our formalism reveals that in voltage-driven transport, the translocating polymer suppresses the induced magnetic field via the {\it steric} blockage of the ion current through the mid-pore. In the case of pressure-driven transport, the magnetic field reduction by translocation originates from the negative {\it electrokinetic} contribution of the anionic DNA surface charges to the streaming current predominantly composed of salt cations. The magnitude of the corresponding field signals is located in the nano-Tesla range covered by the resolution of the magnetoelectric sensors able to detect magnetic fields down to the pico-Tesla range. This suggests that the integration of magnetic field detection techniques into the current biosequencing approaches can complement efficiently the conventional biosensing strategies employing ionic current readouts with high susceptibility to background noise.

\end{abstract}

\date{\today}
\maketitle

\section{Introduction}

Macromolecular transport through nanoscale pores is a widespread phenomenon omnipresent in various {\it in vivo} and {\it in vitro} systems. From the viral infection of cells to nanopore-based biosensing techniques~\cite{gn1,gn2,gn3}, the significance of this universal phenomenon for our understanding and control of a large variety of nanoscale processes has motivated intensive research work on the characterization of the so called {\it driven polymer translocation} through membrane nanopores confining salt solutions~\cite{PolyRev}. 

Polymer translocation is equally considered as a potentially rapid and cheap genetic sequencing technique. The holy grail of this scientific project consists in extracting the genetic sequence of the translocating polymer directly from the ionic current variations caused by the interaction of the macromolecule with the surrounding electrolyte. Driven by this challenging task, during the last two decades, translocation experiments have intensively probed the correlation between the ion currents and the nucleotide structure of polymers crossing biological or synthetic nanopores under externally applied voltages, hydrostatic pressures, and salt concentration gradients~\cite{e1,e2,e3,e4,e5,e6}. 

Owing to the out-of-equilibrium nature of the translocation process, as well as the strong entanglement of the steric, electrostatic, and hydrodynamic interactions governing the dynamics of the polymer-electrolyte complex, the comprehensive characterization of polymer translocation presents itself as a substantially challenging task. The computational investigation of this problem has been initialized with Langevin simulations taking into account solely the polymer conformations and the steric polymer-membrane interactions~\cite{s1,s2}. These numerical studies have characterized the scaling of the mean dwell time with the length of the translocating sequence. Moreover, the non-equilibrium dynamics of polymer translocation has been analytically investigated via a coarse-grained Fokker-Planck approach by Lubensky and Nelson~\cite{th0}. Explicit polymer conformations during translocation have been subsequently taken into account via a tension propagation theory by Sakaue et al.~\cite{th1,th2}.

Due to the predominantly electrostatic nature of the translocation process governed by the strong interactions of the DNA charges, the nanopore surface charges, the salt ions, and the external field in the pore, any predictive theory of polymer translocation should include these electrostatic features in an explicit manner. Driven by this necessity, explicit electrostatics has been incorporated into simulations in Refs.~\cite{s3,s4,s5}. At this point, it is noteworthy that the collective dynamics of the polymer-electrolyte complex is intimately related to the hydrodynamic coupling of the voltage-driven electroosmotic (EO) flow and the surface of the translocating macromolecule. In order to account for this additional complication, Ghosal formulated an ingenious coupling of the Poisson-Boltzmann (PB) and Stokes equations. This led to the first mean-field (MF)-level electrohydrodynamic theory of polymer translocation~\cite{th3}. In addition, Wong and Muthukumar studied polymer capture by EO flows~\cite{th4}, and the non-equilibrium features of polymer translocation via comparison with experiments~\cite{th5,th6}. Finally, in our earlier papers, we developed a unified electrohydrodynamic theory of polymer translocation and ion transport accounting for the electrostatic charge correlations, the coupling of the polymer and membrane charges, and the externally applied hydrostatic pressure gradients inducing streaming currents~\cite{SB2014,SB2018,SB2022}.

\begin{figure}[h!]
\centering
\includegraphics[width=9.2cm]{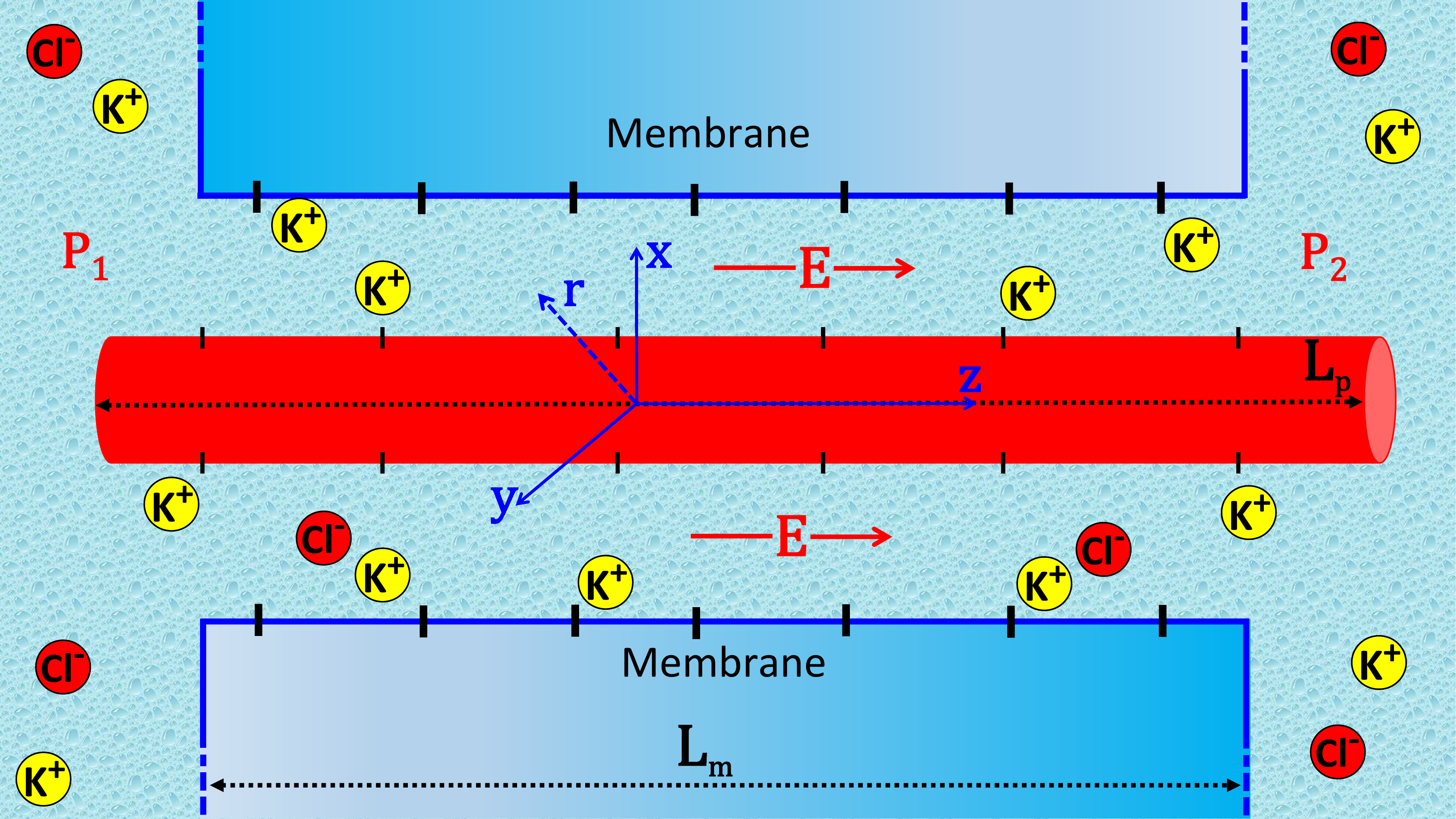}
\includegraphics[width=6.8cm]{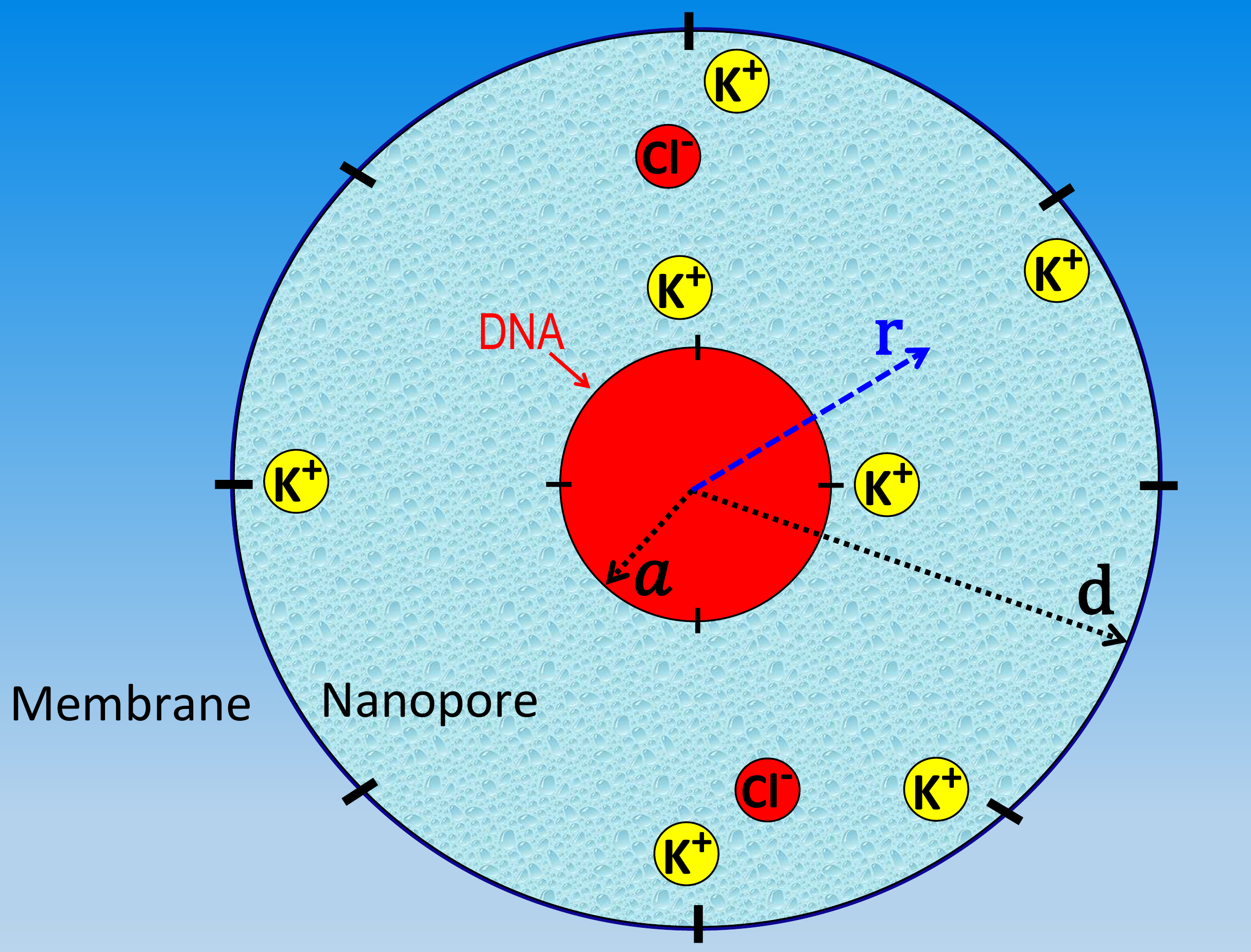}
\caption{Side view (left panel) and cross-section (right panel) of the nanopore with length $L_{\rm m}$ and radius $d$ crossed by the polymer with length $L_{\rm p}$ and radius $a$ under the effect of the uniform electric field $\bE=E\hat{u}_z$ and the hydrostatic pressure gradient $\Delta P=P_1-P_2$. The salt solution KCl has reservoir concentration $\rho_{\rm b}$.}
\label{figure1}
\end{figure}

In the field of nanopore-based polymer translocation, the ongoing technological challenge consists in improving the resolution of the current signals induced by the translocation events via the reduction of the background noise effects. This requirement calls for the integration of alternative polymer sensing techniques complementing the common sequencing approaches employing the ionic current readouts highly sensitive to background noise. Motivated by this need, in this article, we investigate the prospect of detecting polymer translocation events via induced magnetic fields. Ionic current-induced magnetic fields has been previously investigated for polymer-free open pores~\cite{Mag1,Mag2}. Here, by using our electrohydrodynamic theory of ion and polymer transport, we carry-out the first evaluation of the magnetic field signals induced by voltage- and pressure-driven translocation events. We find that the corresponding magnetic field signals in the nano-Tesla range are located within the resolution of the magnetoelectric sensors capable of detecting magnetic fields down to the pico-Tesla range~\cite{pT,nT}.

Our manuscript is organized as follows. Section~\ref{mod} introduces our electrostatic model of voltage- and hydrostatic pressure-driven polymer transport. This model is schematically depicted in Figure~\ref{figure1}. Section~\ref{trform} is devoted to the review of the ion and polymer transport formalism developed in our earlier works of Refs.~\cite{SB2014,SB2018,SB2022}. Then, in Section~\ref{mg}, we use this charge transport theory to calculate the magnetic fields induced by the transport of ions and polyelectrolytes through nanopores. In Section~\ref{res}, within this theoretical framework, we characterize the magnetic field signals triggered by voltage- and pressure-driven DNA translocation events. Our main results are summarized, and the potential extensions of our model are discussed in Conclusions.

\section{Charge transport formalism and induced magnetic fields}

\subsection{Model}
\label{mod}

We introduce here the electrostatic model depicted in Figure~\ref{figure1}. The membrane of thickness $L_{\rm m}$ is placed in an ionic reservoir containing a KCl solution of bulk concentration $\rho_{\rm b}$. The nanopore crossing the membrane along the $z$ axis is a cylindrical hole of radius $d$, length $L_{\rm m}\gg d$, and negative surface charge density $-\sigma_{\rm m}$. The translocating DNA molecule is modeled as a rigid concentric cylinder of radius $a<d$, length $L_{\rm p}\gg d$, and negative surface charge density $-\sigma_{\rm p}$. In the nanopore, the translocation of the polymer captured from the reservoir is either driven by the drift of the externally applied electric field $\bE=E\hat{u}_z$ whose magnitude $E=\Delta V/L_{\rm m}$ is set by the external voltage $\Delta V$, or under the effect of the streaming current induced by the hydrostatic pressure gradient $\Delta P$.

\subsection{Charge transport formalism}
\label{trform}

We review here the charge transport formalism of Refs.~\cite{SB2014,SB2018,SB2022} that will be used to evaluate the magnetic fields induced by nanoconfined ion and polymer transport.

\subsubsection{Computation of the convective liquid velocity}
\label{lvel} 

Our transport formalism is based on the calculation of the convective liquid velocity $u\ce(\br)$ obtained from the solution of the Stokes and Poisson equations,
\bea\label{eq1}
&&\eta\nabla^2 u\ce(\br)+e\bE\rho\ce(\br)-\frac{\Delta P}{\lm}=0;\\
\label{eq2}
&&\nabla^2\phi(\br)+4\pi\ell_{\rm B}\rho\ce(\br)=0,
\eea
The Stokes Eq.~(\ref{eq1}) satisfied by the convective liquid velocity $u\ce(\br)$ includes the viscosity coefficient of water $\eta=8.91\times 10^{-4}\;\mathrm{Pa}\;\mathrm{s}$, the electron charge $e=1.602\times10^{-19}$ C, and the electrolyte charge density 
\be\label{chden}
\rho\ce(\br)=\sum_{i=\pm}q_i\rho_i(\br),
\ee
with the ion number density $\rho_i(\br)$ of the species $i=\pm$ whose dependence on the electrostatic potential will be specified below. Moreover, the Poisson Eq.~(\ref{eq2}) is solved by the average electrostatic potential $\phi(\br)$ induced by the pore and the polymer surface charges. This dimensionless potential is related to the physical electrostatic potential $U(\br)$ as $\phi(\br)=eU(\br)/(k_{\rm B}T)$, with the Boltzmann constant $k_{\rm B}=1.38\times10^{-23}$ J/K, and the liquid temperature $T=300$ K. Eq.~(\ref{eq2}) also contains the Bjerrum length $\ell_{\rm B}=e^2/(4\pi\e_0\ew k_{\rm B}T)$ corresponding to the characteristic interionic distance where the ions interact with the thermal energy $k_{\rm B}T$, with the relative permittivity of water $\ew\approx78$, and the vacuum permittivity $\e_0=8.854\times10^{-12}$ F/m, 

In our model, as the radius of the DNA and the nanopore are much smaller than their respective radii, i.e. $d\ll L_{\rm m}$ and $a\ll L_{\rm p}$, we will neglect edge effects and approximate the nanopore and the polymer as infinitely long concentric cylinders. Owing to the resulting cylindrical symmetry of the system, the liquid velocity, the electrostatic potential, and the ion densities depend exclusively on the radial distance $r$ from the nanopore axis, i.e. $u\ce(\br)=u\ce(r)$, $\phi(\br)=\phi(r)$, and $\rho_i(\br)=\rho_i(r)$.

The hydrodynamic boundary conditions (BCs) that should be imposed to the solution of the Stokes Eq.~(\ref{eq1}) are the no-slip BCs at the pore wall ($r=d$) and on the polymer surface ($r=a$), i.e.
\be
\label{ns}
u\ce(r=d)=0;\hspace{1cm}u\ce(r=a)=v\p,
\ee
where $v\p$ stands for the polymer translocation velocity. Moreover, the electrostatic BCs that should be satisfied by the solution of the Poisson identity~(\ref{eq2}) are the Gauss' laws on the pore and polymer surfaces, i.e.
\be\label{gauss}
\phi'(d)=-4\pi\ell_{\rm B}\sigma_{\rm m};\hspace{1cm}\phi'(a)=4\pi\ell_{\rm B}\sigma_{\rm p}.
\ee

In order to derive the convective liquid velocity and the polymer translocation velocity, first, we substitute the liquid charge density obtained from the Poisson Eq.~(\ref{eq2}) into the Stokes identity~(\ref{eq1}). Expressing the result in the cylindrical coordinate system associated with the symmetry of the nanopore-polymer complex, one obtains
\be
\label{eq3}
\frac{\eta}{r}\partial_rr\partial_r u\ce(r)-\frac{eE}{4\pi\ell_{\rm B}r}\partial_rr\partial_r\phi(r)-\frac{\Delta P}{\lm}=0.
\ee
Integrating Eq.~(\ref{eq3}) twice with respect to the radial coordinate $r$, the convective velocity follows as
\be
\label{eq4}
u\ce(r)=\frac{eE}{4\pi\ell_{\rm B}\eta}\phi(r)+\frac{\Delta P}{4\eta\lm}r^2+c_1\ln r+c_2. 
\ee
In order to fix the integration constants $c_1$ and $c_2$, we impose to Eq.~(\ref{eq4}) the no-slip BCs in Eq.~(\ref{ns}). Moreover, we take into account the steady-state translocation condition set by the force-balance equation $\bF_{\rm el}+\bF_{\rm sh}=\mathbf{0}$ on the polymer, with the electric force $\bF_{\rm el}=-2\pi aL_{\rm p}\sigma_{\rm p}E\hat{u}_z$ and the hydrodynamic shear force $\bF_{\rm sh}=2\pi aL_{\rm p}\eta u\ce'(a)\hat{u}_z$ acting on the polymer surface. Finally, accounting for the second BC in Eq.~(\ref{gauss}), after some algebra, the convective liquid velocity and the polymer translocation velocity follow as
\bea
\label{eq6}
&&u\ce(r)=\frac{eE}{4\pi\ell_{\rm B}\eta}\left[\phi(r)-\phi(d)\right]-\frac{\Delta P}{4\eta\lm}\left[d^2-r^2-2a^2\ln\left(\frac{d}{r}\right)\right],\\
\label{eq7}
&&v\p=\frac{eE}{4\pi\ell_{\rm B}\eta}\left[\phi(a)-\phi(d)\right]-\frac{\Delta P}{4\eta\lm}\left[d^2-a^2-2a^2\ln\left(\frac{d}{a}\right)\right].
\eea

The first and second terms on the r.h.s. of Eqs.~(\ref{eq6})-(\ref{eq7}) are the voltage-driven and pressure-induced velocity components, respectively. In Eq.~(\ref{eq7}), the first term of the voltage-driven velocity $v_{\rm p, EP}=eE\phi(a)/(4\pi\ell_{\rm B}\eta)$ corresponds to the EP component of the polymer mobility. This component oriented along the negative $z$ axis is induced by the direct coupling of the external field $E$ to the anionic DNA surface charges dressed by their counterions (see Figure~\ref{figure1}). Then, the second term $v_{\rm p, EO}=-eE\phi(d)/(4\pi\ell_{\rm B}\eta)$ is the electroosmotically driven polymer mobility component of opposite direction. The latter is induced by the shear force exerted by the diffuse cations brought into the pore by the negatively charged membrane, and driven by the external field $\bE$ along the positive $z$ axis. Finally, the streaming velocity components of Eqs.~(\ref{eq6})-(\ref{eq7}) correspond to a Poiseuille velocity profile augmented by the presence of the cylindrical DNA molecule in the mid-pore.

\subsubsection{Transforming the Poisson Eq.~(\ref{eq2}) into the PB Eq.}

In order to calculate the average potential $\phi(r)$ required for the evaluation of the velocities~(\ref{eq6})-(\ref{eq7}), one has to specify the dependence of the ion density $\rho_i(r)$ in Eqs.~(\ref{eq2})-(\ref{chden}) on the average potential $\phi(r)$. To this aim, we use an electrostatic MF approximation and replace the ion density in Eq.~(\ref{chden}) by the ionic Boltzmann distribution~\cite{el},
\be\label{Bolt}
\rho_i(r)=\rho_{i,{\rm b}}e^{-q_i\phi(r)}, 
\ee
where we introduced the bulk reservoir concentration $\rho_{i,{\rm b}}$ of the ionic species $i$. For the monovalent KCl electrolyte with concentrations $\rho_{i,{\rm b}}=\rho_{\rm b}$ and valencies $q_i=\pm1$, the Poisson Eq.~(\ref{eq2}) takes the form of a second order non-linear differential equation corresponding to the MF-level PB Eq.,
\be\label{eq17}
r^{-1}\partial_rr\partial_r\phi(r)-\kappa^2\sinh\phi(r)=0,
\ee
where we introduced the Debye-H\"{u}ckel screening parameter $\kappa=\sqrt{8\pi\ell_{\rm B}\rho_{\rm b}}$. The PB Eq.~(\ref{eq17}) coupled with the BCs in Eq.~(\ref{gauss}) can be numerically solved with a fourth order Runge-Kutta algorithm.

\subsubsection{Computation of the ion conductance}

The ion current flowing through a cylindrical volume of radius $r$ located between the polymer and the nanopore surfaces ($a<r<d$) is given by the surface integral of the electrolyte flux density,
\be\label{eq8}
i(r)=2\pi e\sum_{i=\pm}q_i\int_a^r\mathrm{d}r'r'\rho_i(r')u_i(r').
\ee
In Eq.~(\ref{eq8}), the net ion velocity $u_i(r)=u\ce(r)+u_{{\rm T},i}$ of the species $i$ is composed of the convective solvent velocity $u\ce(r)$ in Eq.~(\ref{eq6}), and the electrophoretically induced conductive velocity component $u_{{\rm T},i}=\mu_i{\rm sign}(q_i)E$, where  the electrophoretic (EP) ion mobilities are $\mu_+=7.616\times10^{-8}$ $\mbox{m}^2\mbox{V}^{-1}\mbox{s}^{-1}$ for the ${\rm K}^+$ cations, and $\mu_-=7.909\times10^{-8}$ $\mbox{m}^2\mbox{V}^{-1}\mbox{s}^{-1}$ for the ${\rm Cl}^-$ anions~\cite{book}. Expressing Eq.~(\ref{eq8}) in terms of these velocity components, and using Eq.~(\ref{eq6}), the net ion current takes the form of a linear response relation,
\be
\label{eq9}
i(r)=G_V(r)\Delta V+G_P(r)\Delta P,
\ee
with the voltage-driven and streaming conductances of the inner pore zone with radius $a<r<d$,
\bea
\label{eq10}
&&G_V(r)=\frac{2\pi e}{\lm}\sum_{i=\pm}\int_a^r\mathrm{d}r'r'\rho_i(r')\left\{|q_i|\mu_i+\frac{eq_i}{4\pi\ell_{\rm B}\eta}\left[\phi(r')-\phi(d)\right]\right\};\\
\label{eq11}
&&G_P(r)=\frac{\pi e}{2\eta\lm}\int_a^r\mathrm{d}r'r'\rho\ce(r')\left[d^2-r'^2-2a^2\ln\left(\frac{d}{r'}\right)\right].
\eea
The total current $I=i(r=d)$ through the polymer-blocked pore follows from Eq.~(\ref{eq9}) as
\be
\label{eq12}
I=G_V(d)\Delta V+G_P(d)\Delta P.
\ee

From Eqs.~(\ref{eq9})-(\ref{eq11}), the ionic current density $J_i(r)$ of the species $i$ defined as
\be
\label{eqi13}
i(r)=:2\pi\int_a^r{\rm d}r'r'\sum_{i=\pm}J_i(r')
\ee
follows in the form
\be
\label{eqi14}
J_i(r)=\frac{e\Delta V}{L_{\rm m}}\rho_i(r)\left\{|q_i|\mu_i+\frac{eq_i}{4\pi\ell_{\rm B}\eta}\left[\phi(r)-\phi(d)\right]\right\}+\frac{eq_i\Delta P}{4\eta L_{\rm m}}\rho_i(r)\left[d^2-r^2-2a^2\ln\left(\frac{d}{r}\right)\right].
\ee
Eq.~(\ref{eq14}) will allow to identify the local contribution of different salt species to the ionic current. 

We recast now the conductances~(\ref{eq10})-(\ref{eq11})  in a form that facilitates the identification of the current composition. First, we focus on the voltage-driven conductance~(\ref{eq10}). Inserting into the latter equality the identity $\rho\ce(r)=-\partial_r\left[r\partial_r\phi(r)\right]/(4\pi\ell_{\rm B})$ following from the Poisson Eq.~(\ref{eq2}), and performing two successive integrations by parts, one obtains
\bea
\label{eq15}
G_V(r)&=&\frac{2\pi e}{\lm}\sum_{i=\pm}|q_i|\mu_i\int_a^r\mathrm{d}r'r'\rho_i(r')\\
&&+\frac{e^2}{8\pi\ell_{\rm B}^2\lm\eta}\left\{\int_a^r\mathrm{d}r'r'\left[\phi'(r')\right]^2-r\phi'(r)\left[\phi(r)-\phi(d)\right]
-4\pi\ell_{\rm B}\sigma_{\rm p}a\left[\phi(d)-\phi(a)\right]\right\}.\nonumber
\eea
Setting $r=d$, the total voltage-driven conductance follows in the form
\be\label{eq16}
G_V(d)=\frac{2\pi e}{\lm}\sum_{i=\pm}|q_i|\mu_i\int_a^d\mathrm{d}r'r'\rho_i(r')+\frac{e^2}{8\pi\ell_{\rm B}^2\lm\eta}\int_a^d\mathrm{d}r'r'\left[\phi'(r')\right]^2-\frac{e^2\sigma_{\rm p}a}{2\ell_{\rm B}\lm\eta}\left[\phi(d)-\phi(a)\right].
\ee
In Eq.~(\ref{eq16}), the first term is the conductive transport component originating from the electrophoretic mobility of the individual ions. Then, the second and third terms are the convective current contributions originating from the free ions, and the counterions brought by the polymer into the pore, respectively. 

Finally, we consider the streaming conductance~(\ref{eq11}). Eliminating from the latter the ionic charge density with the Poisson Eq.~(\ref{eq2}), and carrying out two integrations by parts, one obtains
\bea
\label{eq13}
G_P(r)&=&-\frac{e}{8\eta\ell_{\rm B}\lm}\left\{\left[d^2-r^2-2a^2\ln\left(\frac{d}{r}\right)\right]r\phi'(r)-4\pi\ell_{\rm B}\sigma_{\rm p}a\left[d^2-a^2-2a^2\ln\left(\frac{d}{a}\right)\right]\right\}\\
&&-\frac{e}{2\eta\ell_{\rm B}\lm}\int_a^r\mathrm{d}r'r'\left[\phi(r)-\phi(r')\right].\nonumber
\eea
Thus, for $r=d$, the total streaming conductance follows as
\be
\label{eq14}
G_P(d)=-\frac{e}{2\eta\ell_{\rm B}\lm}\int_a^d\mathrm{d}r'r'\left[\phi(d)-\phi(r')\right]+\frac{\pi e\sigma_{\rm p}a}{2\eta\lm}\left[d^2-a^2-2a^2\ln\left(\frac{d}{a}\right)\right].
\ee
The first term on the r.h.s. of Eq.~(\ref{eq14}) corresponds to the streaming current component associated with the diffuse ions. Then, the second term accounts for the additional contribution from the counterions bound to the translocating polymer. In the remainder, we will use the identities~(\ref{eq9})-(\ref{eq14}) to evaluate the induced magnetic field signals triggered by the electrophoretically driven and the pressure-induced polymer translocation events. We finally note that the limit of the conductances~(\ref{eq15})-(\ref{eq14}) for DNA-free pores follows from the evaluation of the corresponding identities at $a=0$ and $\sigma_{\rm p}=0$.

\subsection{Derivation of the induced magnetic fields}
\label{mg}

Within the framework of the charge transport formalism developed in the previous part, we derive here the magnetic fields induced by the voltage- and pressure-driven ionic currents during translocation. We consider exclusively the stationary fields and thus neglect any transient effects. Within this static limit, the induced magnetic field $\bb(\br)$ satisfies the Amp\`{e}re law,
\be\label{amp}
\oint_{C(S)} \bb(\br)\cdot{\rm d}{\mathbf\ell}=\mu_0 i_{\rm tot}(r),
\ee
where $C(S)$ is the circular contour of radius $r>a$ enclosing the surface $S=\pi r^2$ parallel with the cross-sectional area of the nanopore~\cite{jackson_book}. Moreover, $\mu_0=4\pi\times10^{-7}$ ${\rm N}/{\rm A}^2$ is the magnetic permeability of the medium, and the function $i_{\rm tot}(r)$ is the ionic current~(\ref{eq9}) augmented by the longitudinally moving DNA surface charges, i.e.
\be
\label{it}
i_{\rm tot}(r)=i(r)-ev_p\oiint_{S}{\rm d}^2S\;\sigma_{\rm p}\delta(r-a)=i(r)-2\pi a\sigma_{\rm p}ev_{\rm p}.
\ee

Due to the cylindrical symmetry of the charges inducing the magnetic field $\bb(\br)$, the latter is purely tangent to the contour of the inner pore surface with radius $r$, and it depends only on the radial distance $r$ from the pore axis. Considering these points in the Amp\`{e}re law~(\ref{amp}), and substituting into the latter identity Eq.~(\ref{it}) together with Eq.~(\ref{eq9}), the induced magnetic field follows as
\be
\label{f1}
B(r)=B_V(r)+B_P(r),
\ee
with the voltage- and pressure-induced field components given by
\bea
\label{f2}
B_V(r)&=&\frac{\mu_0\Delta V}{2\pi r}\left[\bar{G}_V(r)\theta(d-r)+\bar{G}_V(d)\theta(r-d)\right];\\
\label{f3}
B_P(r)&=&\frac{\mu_0\Delta P}{2\pi r}\left[\bar{G}_P(r)\theta(d-r)+\bar{G}_P(d)\theta(r-d)\right].
\eea
In Eqs.~(\ref{f2})-(\ref{f3}), we introduced the ion conductances augmented by the flowing DNA charges taken into account by the second term on the r.h.s. of Eq.~(\ref{it}),
\bea
\label{f4}
\hspace{-5mm}\bar{G}_V(r)&=&\frac{2\pi e}{\lm}\sum_{i=\pm}|q_i|\mu_i\int_a^r\mathrm{d}r'r'\rho_i(r')+\frac{e^2}{8\pi\ell_{\rm B}^2\lm\eta}\left\{\int_a^r\mathrm{d}r'r'\left[\phi'(r')\right]^2-r\phi'(r)\left[\phi(r)-\phi(d)\right]\right\};\\
\label{f5}
\hspace{-5mm}\bar{G}_P(r)&=&-\frac{e}{2\eta\ell_{\rm B}\lm}\int_a^r\mathrm{d}r'r'\left[\phi(r)-\phi(r')\right]-\frac{e}{8\eta\ell_{\rm B}\lm}\left[d^2-r^2-2a^2\ln\left(\frac{d}{r}\right)\right]r\phi'(r).
\eea
Thus, inside the nanopore $r\leq d$, the induced magnetic field components~(\ref{f2})-(\ref{f3}) read
\bea
\label{f6}
B_V(r)&=&\frac{\mu_0\Delta Ve}{\lm r}\sum_{i=\pm}|q_i|\mu_i\int_a^r\mathrm{d}r'r'\rho_i(r')\\
&&+\frac{\mu_0\Delta Ve^2}{\left(4\pi\ell_{\rm B}\right)^2\lm\eta r}\left\{\int_a^r\mathrm{d}r'r'\left[\phi'(r')\right]^2-r\phi'(r)\left[\phi(r)-\phi(d)\right]\right\};\nonumber\\
\label{f7}
B_P(r)&=&-\frac{\mu_0\Delta Pe}{4\pi\eta\ell_{\rm B}\lm r}\int_a^r\mathrm{d}r'r'\left[\phi(r)-\phi(r')\right]-\frac{\mu_0\Delta Pe}{16\pi\eta\ell_{\rm B}\lm}\left[d^2-r^2-2a^2\ln\left(\frac{d}{r}\right)\right]\phi'(r).
\eea

The magnetic field identities~(\ref{f2})-(\ref{f3}) are the main theoretical results of the present work. In the remainder, these equalities will be used to predict the variation of the ionic current-induced magnetic fields by polymer translocation events.

\section{Results}
\label{res}

\begin{figure}[h!]
\centering
\includegraphics[width=17cm]{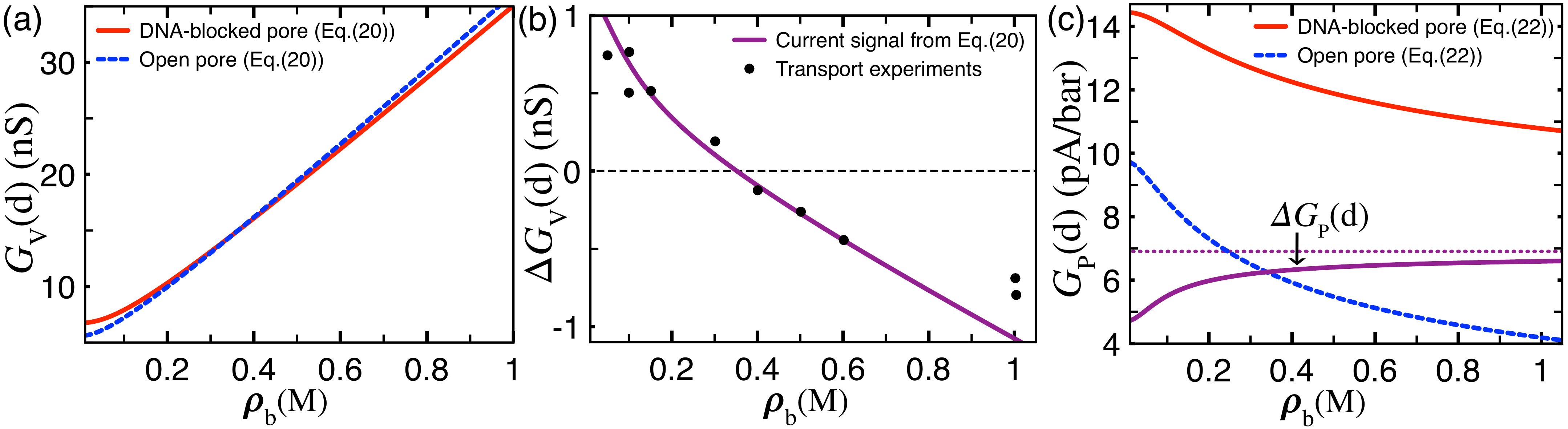}
\caption{(a) Voltage-driven conductance~(\ref{eq16}) in open (blue) and DNA-blocked pores (red), and (b) alteration of the pore conductance by translocation events versus the salt concentration. In (b), the disk symbols are the current blockade data from the transport experiments of Smeets et al.~\cite{e4}, and the solid curve is the theoretical prediction corresponding to the difference of the curves in (a). (c) Streaming conductance~(\ref{eq14}) in open and DNA-blocked pores, and the streaming current signal (solid purple curve) obtained from the difference of these curves. The dotted curve is the DNA-bound counterion contribution corresponding to the second term of Eq.~(\ref{eq14}). See the main text for the model parameters.}
\label{figure2}
\end{figure}

\subsection{Voltage-driven current blockade: theory versus experiments}

In this part, we use the transport formalism introduced in the previous section to probe the ionic current signals induced by the voltage-driven polymer translocation events. Thus, we switch off the pressure gradient and set $\Delta P=0$. The radius of the ds-DNA molecule and its effective smeared surface charge density are $a=1$ nm~\cite{DNAdi} and $\sigma_{\rm p}=0.4$ $\rm{e}/{nm}^2$~\cite{SB2014}. Moreover, the radius and the surface charge of the solid-state nanopore, its thickness, and the external voltage taken from Ref.~\cite{e4} are $d=5$ nm, $\sigma_{\rm m}=0.06$ ${\rm C}/{\rm m}^2$, $L_{\rm m}=34$ nm, and $\Delta V=120$ mV.

Figure~\ref{figure2}(a) displays the conductance of the open and DNA-blocked pores obtained from Eq.~(\ref{eq16}) against the salt concentration. The plot shows that at low concentrations $\rho_{\rm b}\lesssim0.35$ M, the translocated pore exhibits a stronger ion conductance than the open pore. However, at larger salt concentrations $\rho_{\rm b}\gtrsim0.35$ M, the conductance of the DNA-free nanopore exceeds that of the blocked pore. The corresponding current signal obtained from the difference of the blue and red curves is reported in Figure~\ref{figure2}(b) (solid curve). In the low concentration regime, the positive value of the signal ($\Delta G_{\rm V}>0$)  indicates that DNA translocation in dilute salt results in the enhancement of the ionic current through the open pore. Then, upon the rise of the salt concentration from $\rho_{\rm b}\approx0.05$ M to $1.0$ M, the current signal drops, and switches from positive to negative ($\Delta G_{\rm V}<0$)  at $\rho_{\rm b}\approx0.35$ M. Thus, in dense salt solutions, DNA translocation causes the net blockage of the ion current through the nanopore.

In Figure~\ref{figure2}(b), we equally reported the experimental conductance data obtained by Smeets et al.~\cite{e4} (disk symbols). These data have been previously confronted in Ref.~\cite{SB2014} with a correlation-corrected transport theory. Figure~\ref{figure2}(a) displays the first confrontation of the present MF-level transport theory with the experimental data. This comparison shows that in the submolar concentration range, our theory neglecting the charge correlations considered in Ref.~\cite{SB2014} exhibits a good quantitative agreement with the experiments. The deviation of the theoretical prediction from the experimental data at the molar concentration $\rho_{\rm b}=1.0$ M may be due to the onset of significant ionic hard-core interactions absent in the present continuous theory. This said, the overall agreement of our MF formalism with the experimental conductivity indicates that monovalent ion transport and polymer translocation through large solid-state nanopores are mainly governed by MF-level electrohydrodynamics.

\begin{figure}[h!]
\centering
\includegraphics[width=17cm]{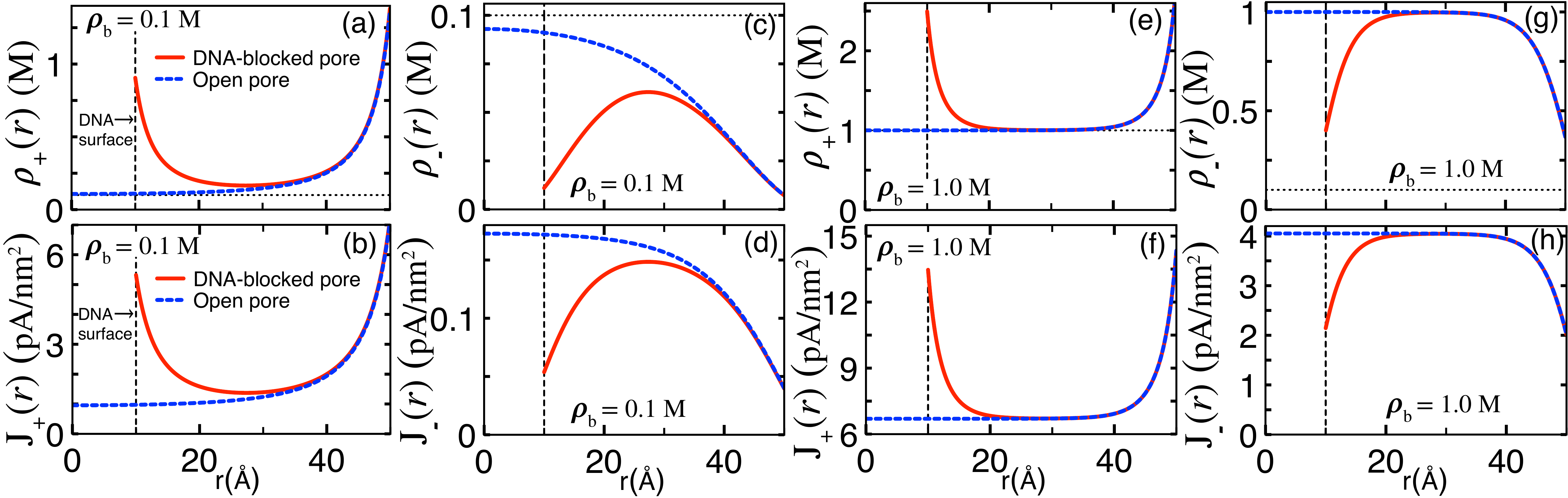}
\caption{Radial density distribution~(\ref{Bolt}) (top plots) and the voltage-driven ionic current density~(\ref{eqi14}) (bottom plots) of the species ${\rm K}^+$ and ${\rm Cl}^-$ in open (dashed blue) and DNA-blocked pores (red) at the salt concentration (a)-(d) $\rho_{\rm b}=0.1$ M and (e)-(h) $\rho_{\rm b}=1.0$ M. The model parameters are the same as in Figures~\ref{figure2}(a)-(b).}
\label{figure3}
\end{figure}

In order to illustrate the mechanism behind the non-uniform effect of the DNA translocation on the ion current, in Figure~\ref{figure3}, we display the radial pore distribution~(\ref{Bolt}) (top plots) and the ionic current density~(\ref{eqi14}) (bottom plots) of the species ${\rm K}^+$ and ${\rm Cl}^-$ in open (dashed blue curves) and DNA-blocked pores (solid red curves) at two different salt concentrations. First, Figures~\ref{figure3}(a)-(d) indicate that in dilute salt ($\rho_{\rm b}=0.1$ M), the main contribution to the ion current through the DNA-free pore originates from the interfacial ${\rm K}^+$ counterions attracted by the anionic pore wall. Then, one notes that DNA penetration into the pore brings two opposing effects to ion transport. Namely, the translocation event blocks the weak ion current through the mid-pore region at $r<10$ {\AA}, but the anionic polymer also brings extra ${\rm K}^+$ counterions flowing in the cylindrical layer $10\;{\rm {\AA}}<r\lesssim20\;{\rm {\AA}}$. As the cumulative current density induced by these additional counterions (red curve) is significantly higher than that associated with the mobile ions blocked by the DNA molecule (blue curve), DNA translocation in dilute salt leads to the net current increment observed in the low concentration regime of Figure~\ref{figure2}(b).

Figures~{\ref{figure3}(e)-(h) indicate that in the opposite regime of large salt concentrations ($\rho_{\rm b}=1.0$ M) where the ion transport in the mid-pore region is governed by bulk-like elektrokinetics, the number and current densities of the ${\rm K}^+$ and ${\rm Cl}^-$ ions in the zone $r<10$ {\AA} of the open pore are comparable with that of the ${\rm K}^+$ cations attracted by the pore wall and the DNA surface. As a result, during translocation, the negative contribution from the DNA-induced ion blockage to the net current dominates the incremental effect of the extra ${\rm K}^+$ counterions dragged by the DNA molecule into the pore. Consequently, DNA translocation in dense salt solutions leads to the net current decrement observed in the large concentration regime of Figure~\ref{figure2}(b). We note that this transport picture emerging from our quantitative analysis agrees with the electroneutrality-based qualitative explanation proposed in Ref.~\cite{e4}.

\subsection{Magnetic field signals induced by voltage-driven currents}

We predict now the magnetic field signals associated with these voltage-driven polymer translocation events. The top panels of Figure~\ref{figure4} display the magnetic field profile in Eq.~(\ref{f6}) at various salt concentrations indicated in the legends. These plots indicate that over the entire concentration range of Figures~\ref{figure2}(a)-(b), the voltage-driven ion current flowing along the positive $z$ axis of the DNA-free pore induces a positive magnetic field of the order $B_{\rm V}(r)\sim10-100$ nT rising monotonically from the pore axis to the pore wall (blue curves).

Figure~\ref{figure4}(a) shows that in the dilute salt regime, DNA translocation modifies the corresponding magnetic field of the open pore at three different levels (red curve). Namely, within the charge-free DNA volume at $r<10$ {\AA}, the polymer cancels the magnetic field, i.e. $B_V(r)=0$. Then, in the immediate vicinity of the DNA molecule dragged by the EO flow along the positive z axis, the anionic DNA surface charges generate a weakly negative magnetic field  of order $B_V(r)\sim-2$ nT, thereby inverting locally the sign of the positive field predominantly induced by the ${\rm K}^+$ cations in the DNA-free nanopore. Finally, moving away from the DNA surface towards the nanopore wall, the ${\rm K}^+$ cations attracted by the negatively charged membrane and the additional ${\rm K}^+$ ions brought by the DNA molecule turn the magnetic field from a negative to a positive value higher the magnetic field of the open pore. Thus, in dilute salt, DNA translocation enhances the current-induced magnetic field of the DNA-free pore.

In Figure~\ref{figure4}(b), we illustrate the radial distribution of the resulting magnetic field signal during translocation. One sees that within the DNA volume as well as in the close vicinity of the DNA surface, the ionic current blockage caused by translocation gives rise to a magnetic field decrement ($\Delta B_V(r)<0$) characterized by a signal minimum $\Delta B_V(a)\sim-4$ nT at the DNA boundary. Outside this mid-pore zone, the increment of the magnetic field caused by the extra ${\rm K}^+$ ions brought by DNA switches the signal from a negative to a positive value of a peak magnitude $\Delta B_V(r)\sim4$ nT.

\begin{figure}[h!]
\centering
\includegraphics[width=17cm]{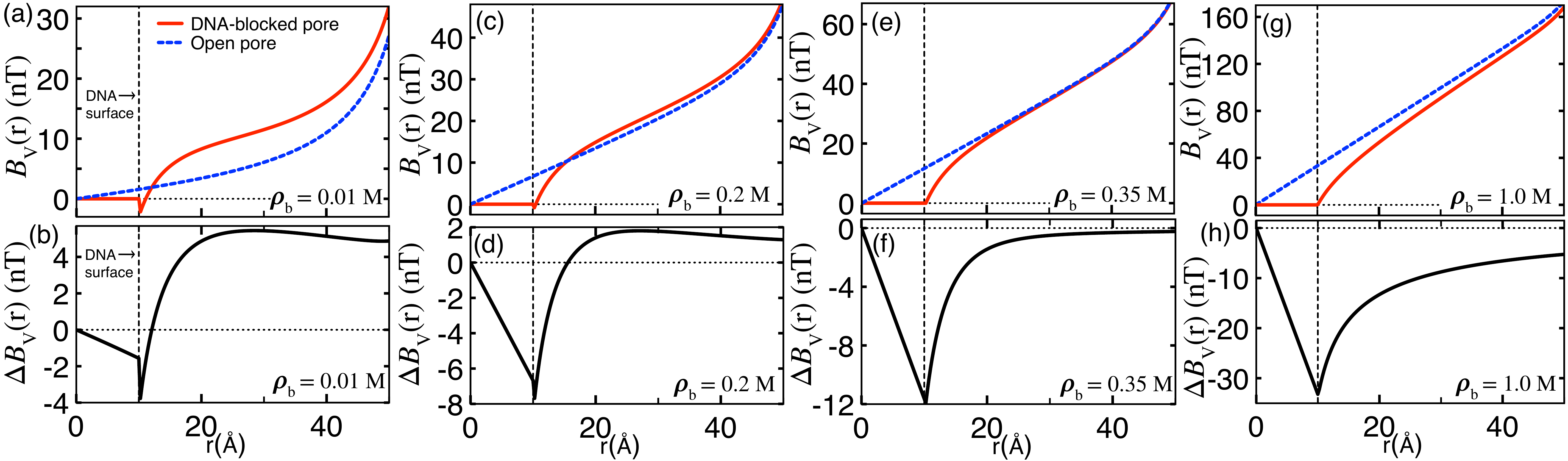}
\caption{Top: Radial profile of the magnetic field~(\ref{f6}) induced by the voltage-driven current in open (blue) and DNA-blocked pores (red) at various salt concentrations indicated in the legends. Bottom: The magnetic field signal corresponding to the difference of the curves in the top plots. The model parameters are the same as in Figures~\ref{figure2}(a)-(b).}
\label{figure4}
\end{figure}

Figures~\ref{figure4}(c), (e), and (g) show that rising the salt concentration beyond the value $\rho_{\rm b}\approx0.2$ M  where the negative contribution of the ionic current blockage by the DNA volume starts to take over the incremental effect of the DNA-bound counterions to the net current, the enhancement of the magnetic field by these counterions during translocation weakens gradually.  As a result, in the high salt concentration regime $\rho_{\rm b}\gtrsim0.35$ M governed by the DNA-induced current blockage,  polymer translocation weakens the magnetic field in the entire pore. The impact of this feature on the field signal is illustrated in Figures~\ref{figure4}(d), (f), and (h). The plots show that upon salt increment, the signal minimum at the DNA surface deepens quickly, and the positive signal branch dissipates gradually. In the salt concentration regime $\rho_{\rm b}\gtrsim0.35$ M, the magnetic field signal becomes purely negative, and the signal minimum drops proportionally with the amount of added salt.

\subsection{Magnetic field signals induced by streaming currents}

In this section, we predict the magnetic field signals induced by the alteration of the streaming currents during pressure-driven polymer translocation events. Thus, we turn off the external voltage ($\Delta V=0$), and set the external pressure to $\Delta P=1$ bar. In Figure~\ref{figure2}(c), we illustrate first the variation of the streaming conductance with the salt concentration. This plot shows that the increment of salt weakens monotonically the positive conductance of DNA-free and blocked pores, i.e. $\rho_{\rm b}\uparrow G_{\rm p}(d)\downarrow$.

In order to explain this feature, we first note that in contrast to the voltage-driven conductance~(\ref{eq10}) largely dominated by the individual bulk conductivities of the ion species ${\rm K}^+$ and ${\rm Cl}^-$, the streaming conductance~(\ref{eq11}) of purely convective origin is set by the net charge density $\rho\ce(r)$ of the electrolyte confined to the pore. As the anionic membrane and DNA surface charges give rise to the ${\rm K}^+$ excess and the ${\rm Cl}^-$ deficiency inside the nanopore,  the resulting charge imbalance leads to a positive charge density ($\rho\ce(r)>0$), and a ${\rm K}^+$-rich streaming conductance of positive sign ($G_P(d)>0$). Then, upon the rise of the salt concentration, the enhancement of the Debye screening experienced by the electrostatic potential in Eq.~(\ref{eq14}) suppresses this charge imbalance, decreasing the electrolyte charge density and the streaming conductance of the nanopore.

We focus now on the streaming current signal. In contrast to the non-uniform alteration of the voltage-driven current by translocation (see Figures~\ref{figure2}(a)-(b)), the comparison of the blue and red curves in Figure~\ref{figure2}(c) shows that the translocating polymer leads to the net enhancement of the streaming conductance in the entire salt concentration regime. For a clear insight into this feature, we also reported the streaming conductance signal (solid purple curve) together with the second term of Eq.~(\ref{eq14}) (dotted horizontal curve) embodying solely the current contribution from the ${\rm K}^+$ cations bound to the polymer. The closeness of these two curves indicates that streaming current amplification by translocation is mainly due to the extra ${\rm K}^+$ cations brought by DNA into the nanopore.

In Figures~\ref{figure5}(a)-(b), we display the radial profile of the streaming conductance~(\ref{eq13}) in open (blue) and DNA-blocked pores (red), and the profile of the corresponding current signal $\Delta G_{\rm p}(r)$ (red) at the dilute salt concentration $\rho_{\rm b}=0.01$ M. In agreement with the conductance curves of Figure~\ref{figure2}(c), these plots show that polymer penetration amplifies the local streaming current, generating a positive current signal $\Delta G_P(r)>0$ at all points in the liquid region $a<r<d$.

\begin{figure}[h!]
\centering
\includegraphics[width=17cm]{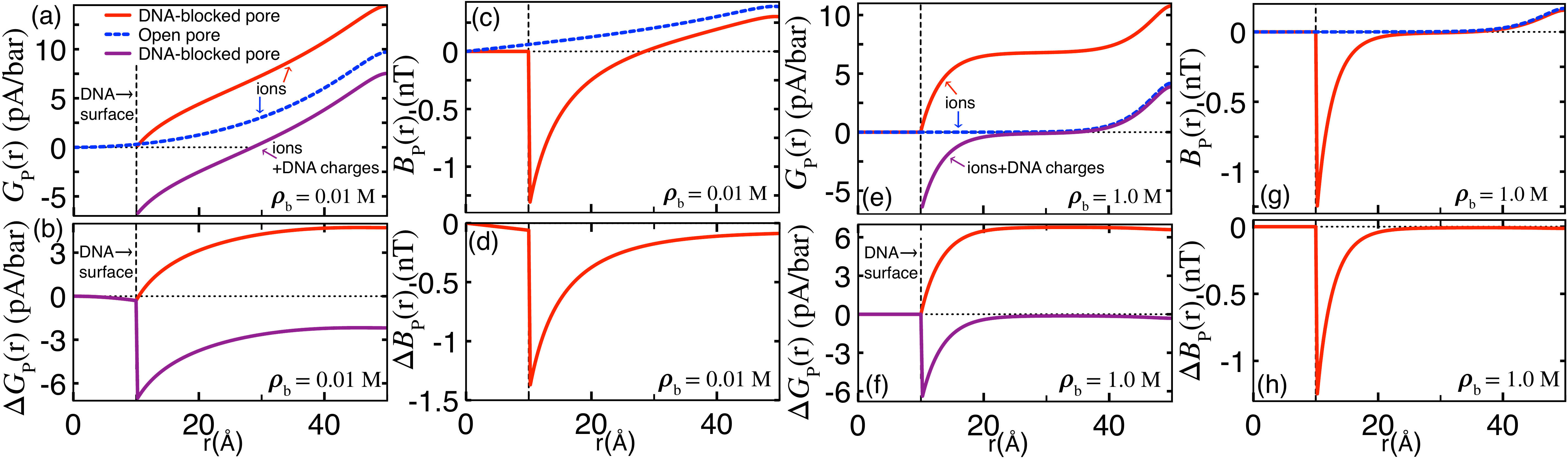}
\caption{(a) Radial profile of the streaming conductance in Eq.~(\ref{eq13}) in open (dashed blue) and DNA-blocked pores (red), and the conductance~(\ref{f5}) augmented by the translocating DNA charges (purple). (b) Variation of the conductance by translocation corresponding to the difference of the curves for open and blocked pores in (a). (c) Radial profile of the magnetic field~(\ref{f7}) induced by the streaming current in open (blue) and DNA-blocked pores (red), and (d) the magnetic field signal corresponding to the difference of these curves. The salt concentration is $\rho_{\rm b}=0.01$ M. The panels (e)-(h) display the plots in (a)-(d) at the molar concentration $\rho_{\rm b}=1.0$ M. The pressure gradient is $\Delta P=1$ bar. The remaining parameters are the same as in Figures~\ref{figure2}(a)-(b).}
\label{figure5}
\end{figure}

Figures~\ref{figure5}(c)-(d) illustrate the radial profile of the corresponding magnetic fields and the field signal. As expected, one sees that the magnetic field induced by the cationic streaming current through the open pore is purely positive. Then, one notes that in the DNA-blocked nanopore, as one moves from the DNA surface to the pore wall, the magnetic field evolves from a strongly negative value to a positive value lower than the field magnitude of the open pore. Thus, despite the streaming current increment originating from the extra ${\rm K}^+$ cations brought by DNA into the pore, translocation gives rise to the reduction and the sign inversion of the magnetic field, leading to a negative field signal $\Delta B_P(r)<0$ whose minimum is sharply localized at the DNA boundary. 

This seemingly counterintuitive effect can be explained by noting that the streaming current-induced magnetic field~(\ref{f7}) is related to the DNA charge-dressed conductance $\bar{G}_P(r)$ in Eq.~(\ref{f5}) rather than the purely ionic conductance $G_P(r)$ in Eq.~(\ref{eq13}). In order to evaluate the additional effect of the DNA charge flow on the total charge current, in Figures~\ref{figure5}(a)-(b), we reported the dressed conductance~(\ref{f5}) and the corresponding current signal (purple curves). Indeed, one sees that while DNA translocation enhances the ionic conductance ($\Delta G_P(r)>0$), the additional contribution from the DNA surface charges leads to a net decrease of the total charge flow in the entire pore ($\Delta \bar{G}_P(r)<0$). This leads in turn to the decrement and the sign inversion of the magnetic field in Figure~\ref{figure5}(c). 

Thus, while the magnetic field signal induced by the voltage-driven translocation stems from the steric blockage of the ion conductivity by the DNA volume in the mid-pore, in Figure~\ref{figure5}(d), the field signal triggered by the pressure-driven translocation is mainly due to the negative magnetic field component induced by the anionic surface charges of the translocating DNA molecule. This effect becomes more obvious at the molar salt concentration $\rho_{\rm b}=1.0$ M of  Figures~\ref{figure5}(e)-(h). Therein, one notes that the variation of the total charge conductance $\Delta\bar{G}_P(r)$ and the induced magnetic field $\Delta B_V(r)$ is indeed solely due to the DNA charges dressed by their counterions at $10\;{\rm {\AA}}< r\lesssim20\;{\rm {\AA}}$. 

Interestingly, the comparison of Figures~\ref{figure5}(c) and (g) indicates that the increment of salt concentration by two orders of magnitude does not affect the depth of the magnetic field in the translocated pore. This is in contrast with the top plots of Figure~\ref{figure4} where one sees that the weak negative field at $r=a$ induced by the voltage-driven DNA decays quickly with the increment of salt. This point can be clarified by inspecting the analytical form of these magnetic field minima obtained from Eqs.~(\ref{f6}) and~(\ref{f7}) as
\bea\label{f10}
B_V(a)&=&-\frac{\mu_0\Delta Ve^2\sigma_{\rm p}}{4\pi\ell_{\rm B}\eta L_{\rm m}}\left[\phi(a)-\phi(d)\right];\\
\label{f11}
B_P(a)&=&-\frac{\mu_0e\sigma_{\rm p}\Delta P}{4\eta L_{\rm m}}\left[d^2-a^2-2a^2\ln\left(\frac{d}{a}\right)\right].
\eea
First, one notes that the voltage-driven magnetic field minimum~(\ref{f10}) is proportional to the average electrostatic potential subject to the Debye screening by added salt. This leads to the salt-driven suppression of the magnetic field minimum observed in the top plots of Figure~\ref{figure4}. However, Eq.~(\ref{f11}) indicates that in pressure-driven translocation, the magnetic field minimum induced solely by the bare DNA charges is independent of the salt concentration. This explains the invariance of the magnetic field at the DNA surface by salt increment from Figures~\ref{figure5}(c) to (g).

\section{Conclusion}

In this work, we reported the first evaluation of the induced magnetic fields during voltage- and pressure-driven translocation events. In order to verify the accuracy of our MF-level transport formalism in predicting the ion currents at the origin of these magnetic fields, we confronted the theoretical ion conductance curves with the experimental conductance data of Ref.~\cite{e4}. In the submolar salt concentration regime, our theory was shown to reproduce the trend of the experimental data with quantitative precision. The agreement of our MF-level transport theory with experiments indicates that electrostatic correlations neglected in the present work have a negligible effect on DNA translocation through large solid-state pores confining monovalent salt solutions.

In the physical conditions and the salt concentration range of these transport experiments, we showed that the voltage-driven current in open pores induces a positive magnetic field of the order $B_V(r)\sim10-100$ nT. During translocation events at dilute salt concentrations $\rho_{\rm b}\lesssim0.35$ M, the steric blockage of this current by DNA reduces the magnetic field, and results in a negative field signal ($\Delta B_V(r)<0$) in the vicinity of the translocating molecule. However, outside this mid-pore region, the extra ${\rm K}^+$ cations brought by the polymer lead to an increment of the magnetic field and induce a positive field signal ($\Delta B_V(r)>0$). Finally, in the large salt concentration regime $\rho_{\rm b}\gtrsim0.35$ M mainly governed by the ionic current blockage by DNA, translocation events reduce the magnetic field in the entire liquid region $a<r<d$.

In pressure-driven DNA translocation events, the additional ${\rm K}^+$ cations dragged by the anionic polyelectrolyte into the pore amplify the streaming current of the open pore at all salt concentrations. However, despite this current enhancement, we found  that pressure-driven translocation suppresses the positive magnetic field of the open pore, and even inverts the sign of this field in the vicinity of the translocating molecule. This peculiarity was shown to originate from the contribution of the longitudinally moving DNA charges to the magnetic field; as these anionic surface charges weaken the effect of the ${\rm K}^+$ cations at the origin of the positive magnetic field, pressure-driven DNA translocation induces a negative field signal ($\Delta B_P(r)<0$) in the entire nanopore and at all experimentally relevant salt concentrations.

Thus, while the weakening of the induced magnetic field caused by voltage-driven translocation is due to the {\it steric} blockage of the ionic current by the DNA volume, the reduction of the magnetic field by pressure-driven translocation originates from the negative {\it electrokinetic} contribution of the anionic DNA surface charges to the ${\rm K}^+$-rich streaming current at the origin of this field. We emphasize that the order of magnitude of the corresponding field signals in the nano-Tesla range is located within the precision of the magnetoelectric sensors able to measure magnetic fields down to the pico-Tesla range~\cite{pT,nT}. Therefore, our characterization of the correlation between the ionic currents and the magnetic field signals during translocation can provide guiding information for the predictive design of nanopore-based biosensing devices integrating magnetic field detection techniques. 

In the present work, we focused exclusively on the magnetic fields induced by monovalent charge transport governed by MF-level electrohydrodynamics. The breaking of this MF picture by the presence of multivalent ions is a relevant complication to be considered in future works. Indeed, in our earlier works on polymer translocation, it was shown that the strong coupling interactions between the multivalent ions, the membrane surface charges, and the polarization charges give rise to exotic electrohydrodynamic behavior, such as the transport of anionic polymers along the external electric field, the voltage-driven transport of overall neutral polymers via dielectrically induced EO flows, and the giant amplification of the ionic current signals by multivalent cations during anionic polymer translocation~\cite{SB2014,SB2018,SB2022}. The prediction of the underlying many-body effects behind these unconventional phenomena on the induced magnetic fields will provide a more complete characterization of the mapping between these fields and ion transport under nanoconfinement.


\newpage
\begin{thebibliography}{99}
\bibitem {gn1} R. Podgornik, H. H. Strey, V. A. Parsegian, \textit{Molecular Interactions in Lipids, DNA and DNA-lipid Complexes, in Gene Therapy: Therapeutic Mechanisms and Strategies}, 209-239, Marcel Dekker, {\bf 2000}, New York.
\bibitem {gn2} H. Yin, R. L. Kanasty, A. Eltoukhy, A. J.  Vegas, J. R. Dorkin, and D. G. Anderson, "{Non-viral vectors for gene-based therapy}",  \href{https://doi.org/10.1038/nrg3763} {{\em Nat. Rev. Genet.} {\bfseries 15} (2014) 541–555}.
\bibitem {gn3} T. J. Thomas and H. A. Tajmirriahi, "{Polyamine-DNA interactions and development of gene delivery vehicles}",  \href{https://doi.org/10.1007/s00726-016-2246-8}  {{\em Amino Acids} {\bfseries 48} (2016) 2423–2431}. 
\bibitem {PolyRev} W. Wanunu, "{Nanopores: A journey towards DNA sequencing}", \href{https://doi.org/10.1016/j.plrev.2012.05.010}{{\em Phys. Life Rev.} {\bfseries 9} (2012) 125-158}. 
\bibitem {e1}  J. J. Kasianowicz, E. Brandin, D. Branton, and D. W. Deamer, "{Characterization of individual polynucleotide molecules using a membrane channel}", \href{https://doi.org/10.1073/pnas.93.24.13770}{{\em Proc. Natl. Acad. Sci. U.S.A} {\bfseries 93} (1996) 13770-13773}. 
\bibitem {e2} S. E. Henrickson, M. Misakian, B. Robertson, and J. J. Kasianowicz, "{Driven DNA transport into an asymmetric nanometer-scale pore}", \href{https://doi.org/10.1103/PhysRevLett.85.3057}{{\em Phys. Rev. Lett.} {\bfseries 14} (2000) 3057-3060}. 
\bibitem {e3} A. Meller, L. Nivon, and D. Branton, "{Voltage-Driven DNA Translocations through a Nanopore}", \href{https://doi.org/10.1103/PhysRevLett.86.3435}{{\em Phys. Rev. Lett.} {\bfseries 86} (2001) 3435-3438}. 
\bibitem {e4} R. M. M. Smeets, U. F. Keyser, D. Krapf, M.-Y. Wue, N. H. Dekker, and C. Dekker, "{Salt dependence of ion transport and DNA translocation through solid-state nanopores}", \href{https://doi.org/10.1021/nl052107w}{{\em Nano Lett.} {\bfseries 6} (2006) 89-95}. 
\bibitem {e5} M. Wanunu, W. Morrison, Y. Rabin, A. Y. Grosberg, and A. Meller, "{Electrostatic focusing of unlabelled DNA into nanoscale pores using a salt gradient}", \href{https://doi.org/10.1038/nnano.2009.379}{{\em Nature Nanotech.} {\bfseries 5} (2010) 160-165}. 
\bibitem {e6} D. P. Hoogerheide, B. Lu, and J. A. Golovchenko, "{Pressure–Voltage Trap for DNA near a Solid-State Nanopore}", \href{https://doi.org/10.1021/nn5025829}{{\em ACS Nano} {\bfseries 8} (2014) 7384–7391}. 
\bibitem {s1} W. Sung and P. J. Park, "{Polymer Translocation through a Pore in a Membrane}",  \href{https://doi.org/10.1103/PhysRevLett.77.783}{{\em Phys. Rev. Lett.} {\bfseries 77} (1996) 783}. 
\bibitem {s2} T. Ikonen, A. Bhattacharya, T. Ala-Nissila, and W. Sung, "{Unifying model of driven polymer translocation}",  \href{https://doi.org/10.1103/PhysRevE.85.051803}{{\em Phys. Rev. E} {\bfseries 85} (2012) 051803}.  
\bibitem {th0} D. K. Lubensky and D. R. Nelson, "{Nonequilibrium dynamics of polymer translocation and straightening}",  \href{https://doi.org/10.1103/PhysRevE.76.021803}{{\em Biophys. J.} {\bfseries77} (1999) 1824-1838}.  
\bibitem {th1} T. Sakaue, "{Nonequilibrium dynamics of polymer translocation and straightening}",  \href{https://doi.org/10.1103/PhysRevE.76.021803}{{\em Phys. Rev. E} {\bfseries76} (2007) 021803}.  
\bibitem {th2} T. Saito and T. Sakaue, "{Dynamical diagram and scaling in polymer driven translocation}",  \href{https://doi.org/10.1140/epje/i2011-11135-3}{{\em Eur. Phys. J. E} {\bfseries34} (2011) 135}. 
\bibitem {s3} B. Luan and A. Aksimentiev, "{Electro-osmotic screening of the DNA charge in a nanopore}",  \href{https://doi.org/10.1103/PhysRevE.78.021912}{{\em Phys. Rev. E} {\bfseries 78} (2008) 021912}.  
\bibitem {s4} B. Luan and A. Aksimentiev, "{Electric and electrophoretic inversion of the DNA charge in multivalent electrolytes}",  \href{https://doi.org/10.1039/B917973A}{{\em Soft Matter} {\bfseries 6} (2010) 243-246}.  
\bibitem {s5} P. Ansalone, M. Chinappi, L. Rondoni, and F. Cecconi, "{Driven diffusion against electrostatic or effective energy barrier across $\alpha$-hemolysin}",  \href{https://doi.org/10.1063/1.4933012}{{\em J. Chem. Phys.} {\bfseries 143} (2017) 154109}.  
\bibitem {th3} S. Ghosal, "{Effect of Salt Concentration on the Electrophoretic Speed of a Polyelectrolyte through a Nanopore}",  \href{https://doi.org/10.1103/PhysRevLett.98.238104}{{\em Phys. Rev. Lett.} {\bfseries98} (2007) 238104}.  
\bibitem {th4} C. T. A. Wong and M. Muthukumar, "{Polymer capture by electro-osmotic flow of oppositely charged nanopores}",  \href{https://doi.org/10.1063/1.2723088}{{\em J. Chem. Phys.} {\bfseries126} (2007) 164903}.  
\bibitem {th5} M. Muthukumar, "{Communication: Charge, diffusion, and mobility of proteins through nanopores}",  \href{https://doi.org/10.1063/1.4894401}{{\em J. Chem. Phys.} {\bfseries141} (2014) 081104}.  
\bibitem {th6} N. A. W. Bell, M. Muthukumar, and U. F. Keyser, "{Translocation frequency of double-stranded DNA through a solid-state nanopore}",  \href{https://doi.org/10.1103/PhysRevE.93.022401}{{\em Phys. Rev. E} {\bfseries93} (2016) 022401}.  
\bibitem{SB2014} S. Buyukdagli and T. Ala-Nissila, "{Controlling Polymer Translocation and Ion Transport via Charge Correlations}", \href{https://doi.org/10.1021/la503327j}{{\em Langmuir} {\bfseries 30} (2014) 12907-12915}.
\bibitem{SB2018} S. Buyukdagli, "{Facilitated polymer capture by charge inverted electroosmotic flow in voltage-driven polymer translocation}'',
\href{https://doi.org/10.1039/C8SM00620B}{{\em Soft Matter} {\bfseries 14} (2018) 3541-3549}.
\bibitem{SB2022} S. Buyukdagli, "{Dielectric manipulation of polymer translocation dynamics in engineered membrane nanopores}",
\href{https://doi.org/10.1021/acs.langmuir.1c02174}{{\em Langmuir} {\bfseries 38} (2022) 122-131}.
\bibitem {Mag1}  A. Riad, B. Khorshidi, A. Mansouri, and M. Sadrzadeh, "{Transient electroosmotic-driven ionic current magnetic fields in a charged nano-capillary}", \href{https://doi.org/10.1016/j.colsurfa.2021.127374}{{\em Colloids Surf. A Physicochem. Eng. Asp.} {\bfseries 628} (2021) 127374}.
\bibitem {Mag2}  S. A. Tabatabaei, A. Mansouri, A. Tarokh, and S. F. Chini, "{Ionic current magnetic fields in 3D finite-length nanopores and nanoslits}", \href{https://doi.org/10.1140/epjp/s13360-022-02519-8}{{\em Eur. Phys. J. Plus} {\bfseries 137} (2022) 312}.
\bibitem {pT} J. Zhai, Z. Xing, S. Dong, J. Li, and D. Viehland, "{Detection of pico-Tesla magnetic fields using magneto-electric sensors at room temperature}", \href{https://doi.org/10.1063/1.2172706}{{\em Appl. Phys. Lett. } {\bfseries 88} (2006) 062510}.
\bibitem {nT} N. Haned and M. Missous, "{Nano-tesla magnetic field magnetometry using an InGaAs-AlGaAs-GaAs 2DEG Hall sensor}", \href{https://doi.org/10.1016/S0924-4247(02)00386-2}{{\em Sensors Actuators A Phys.} {\bfseries 102}(2003) 216-222}.
\bibitem {el} J. Israelachvili, \textit{Intermolecular and Surface Forces}, {\bf 1992}, Academic Press.
\bibitem{book} R. D. Lide, "{Handbook of Chemistry and Physics}'', 93th edition, \textbf{2012}, CRC Press.
\bibitem {jackson_book} J.D. Jackson, \textit{Classical Electrodynamics}, {\bf 1975}, Wiley 2nd ed., New York.
\bibitem {DNAdi}  A. Rich, A. Nordheim, and A. H.-F. Wang, "{The Chemistry and Biology of Left-Handed Z-DNA}", \href{https://doi.org/10.1146/annurev.bi.53.070184.004043}{{\em Ann. Rev. Biochem.} {\bfseries 53} (1984) 791-846}.
\end{thebibliography}
\end{document}